\documentclass[aps,prd,twoside,twocolumn,nofootinbib,showpacs,floatfix]{revtex4-1}
\usepackage{amssymb}
\usepackage{amsmath,amssymb}
\usepackage{graphicx,bm}
\usepackage{slashed}
\usepackage{times}
\usepackage{epstopdf}
\usepackage{ulem} 
\usepackage[usenames]{color}
\usepackage{float}
\usepackage{subfigure}
\usepackage{subfigure}
\usepackage{rotating}
\usepackage{color}
\usepackage{multirow}
\usepackage{dcolumn}
\usepackage{overpic}
\usepackage{booktabs}
\usepackage{makecell}
\usepackage{diagbox}

\renewcommand\sout{\bgroup \color{red} \ULdepth=-.5ex \ULset}

\usepackage[colorlinks, citecolor=blue,anchorcolor=red,menucolor=red,linkcolor=blue,filecolor=red,runcolor=red,urlcolor=blue,frenchlinks=red]{hyperref}
\begin{document}
\title{A comparison between the $P_c$ and $P_{cs}$ systems}
\author{Kan Chen$^{1}$}\email{chenk$_$10@pku.edu.cn}
\author{Zi-Yang Lin$^{1}$}\email{lzy$_$15@pku.edu.cn}
\author{Shi-Lin Zhu$^{1}$}\email{zhusl@pku.edu.cn}

\affiliation{ $^1$School of Physics and Center of High Energy
Physics, Peking University, Beijing 100871, China}

\begin{abstract}
We construct the effective potentials of the $P_c$ and $P_{cs}$
states based on the SU(3)$_{\text{f}}$ symmetry and heavy quark
symmetry. Then we perform the coupled-channel analysis of the lowest
isospin $P_c$ and $P_{cs}$ systems. The coupled-channel effects play
different roles in the $P_c$ and $P_{cs}$ systems. In the $P_c$
systems, this effect gives minor corrections to the masses of the
$P_c$ states. In the $P_{cs}$ system, the
$\Lambda_c\bar{D}_s-\Xi_c\bar{D}$ coupling will shift the mass of the
$P_{cs}(4338)$ close to the $\Xi_c\bar{D}$ threshold. The
$\Lambda_c\bar{D}^{(*)}_s-\Xi_c\bar{D}^{(*)}$ coupling will also
produce extra $P_{cs}$ states. We discuss the correspondence between
the $P_c$ and $P_{cs}$ states. Our results prefer that the SU(3)
partners of the observed $P_{c}(4312)$, $P_{c}(4440)$, and
$P_{c}(4457)$ in the $P_{cs}$ system have not been found yet.
\end{abstract}

\maketitle

\vspace{2cm}

\section{Introduction}\label{sec1}

Very recently, the LHCb Collaboration announced the observation of a
$P_{cs}(4338)$ signal from the $J/\Psi\Lambda$ mass spectrum in the
$B^-\rightarrow J/\Psi \Lambda \bar{p}$ process \cite{LHCb:2022jad}.
The mass and width of this new pentaquark candidate were measured to
be
\begin{eqnarray}
M_{P_{cs}}&=&4338.2\pm 0.7\pm 0.4\quad \text{MeV},\\
\Gamma_{P_{cs}}&=&7.0\pm 1.2\pm 1.3\quad \text{MeV}.
\end{eqnarray}
Meanwhile, the amplitude analysis prefers the $\frac{1}{2}^-$
spin-parity quantum numbers. The central value of the mass of
$P_{cs}(4338)$ is above the $\Xi_c\bar{D}$ threshold. Thus, this
state can not be directly assigned as the $\Xi_c\bar{D}$ molecular
state. However, the authors of Ref. \cite{Meng:2022wgl} pointed out
that the lineshape of this resonance could be distorted from the
conventional Breit-Wigner distribution if it lies very close to and
strongly couples to the threshold.

Besides the newly observed $P_{cs}(4338)$, the $P_{cs}(4459)$ was
observed at LHCb \cite{LHCb:2020jpq} as a candidate of a
$\Xi_c\bar{D}^*$ molecular state, which agrees well with the
prediction from the chiral effective field theory in Ref.
\cite{Wang:2019nvm}. The strange hidden-charm states were also
discussed in Refs.
\cite{Peng:2020hql,Chen:2020kco,Liu:2020hcv,Zhu:2021lhd,Xiao:2021rgp,Wu:2010jy,Santopinto:2016pkp,Shen:2019evi,Xiao:2019gjd}
and reviewed extensively in Refs. \cite{Chen:2016qju,Guo:2017jvc,Liu:2019zoy,Meng:2022ozq,Chen:2022asf,Dong:2021juy}.

The mass of the $P_{cs}(4459)$ is about 19 MeV below the
$\Xi_c\bar{D}$ threshold. In Ref. \cite{Burns:2022uha}, the author
argued that from heavy quark symmetry, the
$[\Xi_{c}\bar{D}]^{1/2^-}$, $[\Xi_c\bar{D}^*]^{1/2^-}$, and
$[\Xi_c\bar{D}^*]^{3/2^-}$ channels should share identical
potentials and have comparable binding energies. However, the heavy
quark symmetry is drastically violated in the charm system due to
the rather small charm quark mass. With the assignment of the
$P_{cs}(4338)$ and $P_{cs}(4459)$ as the $[\Xi_{c}\bar{D}]^{1/2^-}$
and $[\Xi_c\bar{D}^*]^{1/2^-}$ ($[\Xi_c\bar{D}^*]^{3/2^-}$)
molecular states, the degeneracy of the $[\Xi_c\bar{D}^*]^{1/2^-}$
and $[\Xi_c\bar{D}^*]^{3/2^-}$ channels is removed by the
couple-channel effects and recoil corrections.

Another novel phenomenon from the $M_{J/\Psi\Lambda}$ invariant
spectrum \cite{LHCb:2022jad} is that there seems to be a structure
around $M=4254$ MeV. To understand this signal, the LHCb checked the
$m(J/\Psi \Lambda)$ distribution close to the $\Lambda_c^+D_s^-$
threshold and found that this signal is not statistically
significant. Nevertheless, the authors in Ref.
\cite{Nakamura:2022jpd} investigated the $P_{cs}(4338)$ and
$P_{cs}(4255)$ pole positions from a unitary
$\Xi_c\bar{D}-\Lambda_c\bar{D}_s$ coupled-channel scattering
amplitude. Besides, the $P_{cs}(4255)$ pole was also found in a
model with the coupling between the meson-baryon molecule and the
compact five-quark state \cite{Giachino:2022pws}. The
$P_{cs}(4255)$ state was also suggested in an effective field theory
framework \cite{Yan:2022wuz}.

The analogy between the observed ($P_c(4312)$, $P_c(4440)$,
$P_c(4457)$) \cite{LHCb:2015yax,LHCb:2019kea} and ($P_{cs}(4338)$,
$P_{cs}$(4459)) states are discussed in Ref.
\cite{Karliner:2022erb,Wang:2022mxy,Wang:2022neq}. However, since
the $\Sigma_c$ and $\Xi_c$ belong to different SU(3)$_{\text{f}}$
multiplets, the relations between the discussed $P_c$ and $P_{cs}$
states are not clear. Besides, the $P_{\Psi ss}^{N}$ pentaquark
states as the partners of the $P_c$ and $P_{cs}$ states are
investigated in Ref. \cite{Ortega:2022uyu}.

If the $P_{cs}$ states and $P_c$ states can be related via
SU(3)$_{\text{f}}$ symmetry, it is important to investigate the
similarities and differences between these two sets of molecular
candidates. In Ref. \cite{Chen:2021spf,Chen:2021cfl}, we discussed
the symmetry properties of different heavy flavor molecular systems
via a quark level Lagrangian. We proposed that the interactions of
different heavy flavor molecules can be related via a generalized
flavor-spin symmetry \cite{Chen:2021spf}. This framework provides a
suitable tool to discuss the similarities between the $P_c$ and
$P_{cs}$ states.

We also notice an important difference between the $P_c$ and
$P_{cs}$ states. The minimal quark components of the $P_c$ and
$P_{cs}$ states are $c\bar{c}nnn$ and $c\bar{c}nns$ ($n=u,d$),
respectively. For the charmed/charmed-strange mesons and baryons,
the SU(3)$_{\text{f}}$ symmetry breaking effects are reflected on
their physical masses, and we need to distinguish the $s$ quark from
$u$, $d$ quarks when we study the $P_{cs}$ systems. Unlike the $P_c$
pentaquarks, the $P_{cs}$ states can couple to two sets of channels,
i.e., the $cns$-$\bar{c}n$ type and $cnn$-$\bar{c}s$ type channels.
In Table \ref{thresholds}, we list the possible open-charm channels
and their thresholds for the $P_c$ and $P_{cs}$ systems.

\begin{table}[htbp]
\renewcommand\arraystretch{1.5}
\caption{The thresholds of the meson-baryon channels associated with
the $P_c$ and $P_{cs}$ systems, we adopt the isospin averaged masses
for the ground charmed mesons and baryons
\cite{ParticleDataGroup:2020ssz}. All values are in units of MeV.}
\begin{tabular}{cc|ccccccccccccc}
\toprule[0.8pt]
\multicolumn{2}{c}{$P_c$}&\multicolumn{4}{c}{$P_{cs}$}\\
\hline
$\Lambda_c\bar{D}$&4153.7&$\Lambda_c\bar{D}_s$&4255.5&$\Xi_c\bar{D}$&4336.7\\
$\Lambda_c\bar{D}^*$&4295.0&$\Lambda_c\bar{D}_s^*$&4398.7&$\Xi_c\bar{D}^*$&4478.0\\
$\Sigma_c\bar{D}$&4320.8&$\Sigma_c\bar{D}_s$&4422.5&$\Xi_c^\prime\bar{D}$&4446.0\\
$\Sigma_c^*\bar{D}$&4385.4&$\Sigma_c^*\bar{D}_s$&4487.1&$\Xi_c^*\bar{D}$&4513.2\\
$\Sigma_c\bar{D}^*$&4462.1&$\Sigma_c\bar{D}_s^*$&4565.7&$\Xi_c^\prime\bar{D}^*$&4587.4\\
$\Sigma_c^*\bar{D}^*$&4526.7&$\Sigma_c^*\bar{D}_s^*$&4630.3&$\Xi_c^*\bar{D}^*$&4654.5\\
\bottomrule[0.8pt]
\end{tabular}\label{thresholds}
\end{table}

In this work, we will take the $P_{cs}(4338)$ as a molecular
candidate and discuss the following three issues:
\begin{itemize}
\item[(1)] Can we understand the minor binding energy of the $P_{cs}(4338)$ (close to the $\Xi_c\bar{D}$ threshold) through a $\Xi_c \bar{D}$-$\Lambda_c\bar{D}_s$ coupled-channel effect?
\item[(2)] Can we produce a $P_{cs}(4254)$ bound state by including the $\Xi_c \bar{D}$-$\Lambda_c\bar{D}_s$ coupled-channel effect with the potential constrained from SU(3)$_{\text{f}}$ symmetry?
\item[(3)] What is the correspondence between the $P_c$ and $P_{cs}$ states if the interactions of the $P_c$ and $P_{cs}$ states obey a generalized flavor-spin
symmetry?
\end{itemize}

This paper is organized as follows. We present our theoretical
framework in Sec.~\ref{sec2} and the corresponding numerical results
and discussions in Sec.~\ref{sec3}. Sec.~\ref{sec4} is the summary.

\section{Framework}\label{sec2}

In Ref. \cite{Chen:2021cfl}, we proposed an isospin criterion and
pointed out that the $P_c$ and $P_{cs}$ states with the lowest
isospin numbers are more likely to form bound states. Based on the
same Lagrangian, we only focus on the $P_c$ and $P_{cs}$ states with
isospin numbers $I=1/2$ and 0, respectively. Thus, we will not
include the $\Sigma^{(*)}_c\bar{D}^{(*)}_s$ channels listed in Table
\ref{thresholds} for the $P_{cs}$ system.

For the $I=1/2$ $P_c$ states, we consider the following channels for
the $J=1/2$ and $3/2$ states
\begin{eqnarray}
J=\frac{1}{2}&:& \Lambda_c\bar{D},\Lambda_c\bar{D}^*,\Sigma_c\bar{D},\Sigma_c\bar{D}^*,\Sigma_c^*\bar{D}^*,\\
J=\frac{3}{2}&:&\Lambda_c\bar{D}^*,\Sigma_c^*\bar{D},\Sigma_c\bar{D}^*,\Sigma_c^*\bar{D}^*.
\end{eqnarray}
Similarly, for the $I=0$ $P_{cs}$ states, we include the following
channels for the $J=1/2$ and $3/2$ states
\begin{eqnarray}
J=\frac{1}{2}&:&\Lambda_c\bar{D}_s,\Lambda_c\bar{D}^*_s,\Xi_c\bar{D},\Xi_c\bar{D}^*,\Xi_c^\prime\bar{D},\Xi_c^\prime\bar{D}^*,\Xi_c^*\bar{D}^*,\label{Pcs12}\\
J=\frac{3}{2}&:&\Lambda_c\bar{D}_s^*,\Xi_c\bar{D}^*,\Xi_c^*\bar{D},\Xi_c^\prime\bar{D}^*,\Xi_c^*\bar{D}^*.\label{Pcs32}
\end{eqnarray}
The result of the $P_c$ ($P_{cs}$) state with $J=5/2$ can be
obtained from a single-channel calculation and was predicted in Ref.
\cite{Chen:2021cfl} in the same framework. Thus, we will not discuss
them further in this work.

\subsection{Lagrangians for the baryon-meson systems}

To describe the $S$-wave interactions between the ground
charmed/charmed-strange baryons and mesons, we introduce the
following quark-level Lagrangian
\cite{Meng:2019nzy,Wang:2019nvm,Wang:2020dhf,Chen:2021cfl}
\begin{eqnarray}
\mathcal{L}=g_s\bar{q}\mathcal{S}q+g_a\bar{q}\gamma_\mu\gamma^5\mathcal{A}^\mu
q.\label{lag}
\end{eqnarray}
Here, $q=(u,d,s)$, $g_s$ and $g_a$ are two independent coupling
constants that describe the interactions from the exchanges of the
scalar and axial-vector meson currents. They encode the
nonperturbative low energy dynamics of the considered heavy flavor
meson-baryon systems.

From this Lagrangian, the effective potential of the light
quark-quark interactions reads
\begin{eqnarray}
\mathcal{V}&=&\tilde{g}_s\bm{\lambda}_1\cdot\bm{\lambda}_2+\tilde{g}_a\bm{\lambda}_1\cdot\bm{\lambda}_2\bm{\sigma}_1\cdot\bm{\sigma}_2.\label{op}
\end{eqnarray}
Here,
\begin{eqnarray}
\bm{\lambda}_1\cdot\bm{\lambda}_2=\lambda_1^8\lambda_2^8+\lambda_1^i\lambda_2^i+\lambda_1^j\lambda_2^j,
\end{eqnarray}
$i$ and $j$ sum from 1 to 3 and 4 to 7, respectively. The operators
$\lambda_1^8\lambda_2^8$ ($\lambda_1^8\lambda_2^8
(\bm{\sigma}_1\cdot\bm{\sigma}_2)$), $\lambda_1^i\lambda_2^i$
($\lambda_1^i\lambda_2^i (\bm{\sigma}_1\cdot\bm{\sigma}_2)$), and
$\lambda_1^j\lambda_2^j$ ($\lambda_1^j\lambda_2^j
(\bm{\sigma}_1\cdot\bm{\sigma}_2)$) arise from the exchanges of the
isospin singlet, triplet, and two doublets light scalar
(axial-vector) meson currents, respectively. The redefined coupling
constants are $\tilde{g}_s\equiv g_s^2/m^2_{\mathcal{S}}$ and
$\tilde{g}_a\equiv g_a^2/m^2_{\mathcal{A}}$.


The Lagrangian in Eq. (\ref{lag}) allows the exchanges of two types
of scalar and axial-vector mesons that have quantum numbers
$I(J^P)=0(0^+)$, $1(0^+)$, $1/2(0^+)$ and $I(J^P)=0(1^+)$, $1(1^+)$,
$1/2(1^+)$, respectively. At present, we can not specifically pin
down the coupling parameter of each exchanged meson in the above six
meson currents. Alternatively, since the mesons in each meson
current have identical interacting Lorentz structure, we use the
coupling constant $\tilde{g}_s$($\tilde{g}_a$) to collectively
absorb the total dynamical effects from the exchange of each scalar
(axial-vector) meson current. In addition, the couplings
$\tilde{g}_s$ ($\tilde{g}_a$) for the scalar (axial-vector) meson
currents with different isospin numbers are the same in the SU(3)
limit.

The effective potential between the $i$-th baryon-meson channel
$B_iM_i$ and the $j$-th baryon-meson channel $B_jM_j$ with total
isospin $I$ and total angular momentum $J$ can be calculated as
\begin{eqnarray}
v_{ij}&=&\left\langle[B_iM_i]_J^I|\mathcal{V}|[B_jM_j]_J^I\right\rangle.
\end{eqnarray}
Here, the $|[B_iM_i]_J^I\rangle$ is the quark-level flavor-spin wave
function of the considered $i$-th channel baryon-meson system
\begin{eqnarray}
|[B_iM_i]_J^I\rangle&=&\sum_{m_{I_1},m_{I_2}}C_{I_1,m_{I_1};I_2,m_{I_2}}^{I,I_z}\phi_{I_1,m_{I_1}}^{B_{if}}\phi_{I_2,m_{I_2}}^{M_{if}}\nonumber\\
&&\otimes\sum_{m_{S_1},m_{S_2}}C^{J,J_z}_{S_1,m_{S_1};S_2,m_{S_2}}\phi_{S_1,m_{S_1}}^{B_{is}}\phi_{S_2,m_{S_2}}^{M_{is}}.\nonumber\\
\label{BMWF}
\end{eqnarray}
In Eq. (\ref{BMWF}), the $\phi_{S_1,m_{S_1}}^{B_{is}}$ and
$\phi_{S_2,m_{S_2}}^{M_{is}}$ are the spin wave functions of the
baryon and meson, respectively. The total spin wave function can be
obtained with the help of SU(2) CG coefficient
$C_{S_1,m_{S_1};S_2,m_{S_2}}^{J,J_z}$. For the flavor wave functions
of the considered baryons ($\phi_{I_1,m_{I_1}}^{B_{if}}$) and mesons
($\phi_{I_1,m_{I_1}}^{M_{if}}$), their explicit forms have been
given in Ref. \cite{Chen:2021spf}. When constructing the total
flavor wave functions of the considered baryon-meson systems, we use
the SU(2) CG coefficient and take the $s$ quark as a flavor singlet.

The coupled-channel Lippmann-Schwinger equation (LSE) reads
\begin{eqnarray}
\mathbb{T}\left(E\right)&=&\mathbb{V}+\mathbb{V}\mathbb{G}\left(E\right)\mathbb{T}\left(E\right),\label{LSE}
\end{eqnarray}
with
\begin{eqnarray}
\mathbb{V}=\left(
  \begin{array}{ccccc}
    v_{11} & \cdots & v_{1i} & \cdots & v_{1n} \\
    \vdots &  & \vdots &  & \vdots \\
    v_{j1} & \cdots & v_{ji} & \cdots & v_{jn} \\
    \vdots &  & \vdots &  & \vdots \\
    v_{n1} & \cdots & v_{ni} & \cdots & v_{nn} \\
  \end{array}
\right),
\end{eqnarray}
\begin{eqnarray}
\mathbb{T}(E)=\left(
  \begin{array}{ccccc}
    t_{11}(E) & \cdots & t_{1i}(E) & \cdots & t_{1n}(E) \\
    \vdots &  & \vdots &  & \vdots \\
    t_{j1}(E) & \cdots &t_{ji}(E) & \cdots & t_{jn}(E) \\
    \vdots &  & \vdots &  & \vdots \\
    t_{n1}(E) & \cdots & t_{ni}(E) & \cdots & t_{nn}(E) \\
  \end{array}
\right),
\end{eqnarray}
and
\begin{eqnarray}
\mathbb{G}(E)=\text{diag}\left\{G_{1}(E),\cdots,G_{i}(E),\cdots,G_{n}(E)\right\}.
\end{eqnarray}
Here,
\begin{eqnarray}
G_{i}=\frac{1}{2\pi^2}\int dq \frac{q^2}{E-\sqrt{m_{i1}^2+q^2}-\sqrt{m_{i2}+q^2}}u^2(\Lambda).\nonumber\\
\end{eqnarray}
The $m_{i1}$ and $m_{i2}$ are the masses of the baryon and meson in
the $i-$th channel, respectively. In our previous work
\cite{Chen:2021spf,Chen:2021cfl}, we use a step function to exclude
the contributions from higher momenta to perform the single-channel
calculation. In the coupled-channel case, we need to further
suppress the contributions from the channels that are far away from
the thresholds of the considered channels. Thus, we introduce a
dipole form factor $u(\Lambda)=(1+q^2/\Lambda^2)^{-2}$ with regular
parameter $\Lambda=1.0$ GeV
\cite{Nakamura:2022jpd,Leinweber:2003dg,Wang:2007iw}.

The pole position of Eq. (\ref{LSE}) satisfies
$||\bm{1}-\mathbb{V}\mathbb{G}||=0$. For the bound state below the lowest channel, we search the bound state solution in the first Riemann sheet of the lowest channel. For the quasi-bound state between the thresholds of the $i$-th and $j$-th channels, we adopt the complex scaling method and replace the
integration variable $q$ by $q\rightarrow q\times \text{exp}(-i
\theta)$ and maintain $0<\theta<\pi/2$ to find the quasi-bound state solution in the first Riemann sheet of the higher $j$-th channel and the second Riemann sheet of the lower $i$-th channel.
\cite{Liu:2016wxq}.

\section{Numerical results}\label{sec3}
\subsection{Determination of $\tilde{g}_s$ and $\tilde{g}_a$}

We first determine the parameters $\tilde{g}_s$ and $\tilde{g}_a$ in
our model. We collect the matrix elements of
$\langle\bm{\lambda}_1\cdot\bm{\lambda}_2\rangle$,
$\langle\bm{\lambda}_1\cdot\bm{\lambda}_2\bm{\sigma}_1\cdot\bm{\sigma}_2\rangle$
for the $P_c$ and $P_{cs}$ states in Tables \ref{PcEF} and
\ref{PcsEF}, respectively.
\begin{table*}[htbp]
\renewcommand\arraystretch{1.5}
\caption{The matrix elements of
$[\langle\bm{\lambda}_1\cdot\bm{\lambda}_2\rangle$,
$\langle\bm{\lambda}_1\cdot\bm{\lambda}_2\bm{\sigma}_1\cdot\bm{\sigma}_2\rangle]$
for the meson-baryon channels associated with the $J^{P}=1/2^-$ and
$3/2^{-}$ $P_c$ systems.}\label{PcEF}
\begin{tabular}{cccccc|cccccccccccccccccccc}
\toprule[0.8pt]
\multicolumn{6}{c|}{$\mathbb{V}_{1/2}^{P_c}$}&\multicolumn{5}{c}{$\mathbb{V}_{3/2}^{P_c}$}\\
\hline
Channel&  $\Lambda_c\bar{D}$&$\Lambda_c\bar{D}^*$ &$\Sigma_c\bar{D}$&$\Sigma_c\bar{D}^*$&$\Sigma_c^*\bar{D}^*$&Channel&  $\Lambda_c\bar{D}^*$&$\Sigma_c^*\bar{D}$&$\Sigma_c\bar{D}^*$&$\Sigma_c^*\bar{D}^*$\\
\hline
$\Lambda_c\bar{D}$&[$\frac{2}{3}$,0]&[0,0]&[0,0]                &[0,$2\sqrt{3}$]             &[0,$2\sqrt{6}$]&$\Lambda_c\bar{D}^*$&[$\frac{2}{3}$,0]&[0,$-2\sqrt{3}$]&[0,2]&[0,$2\sqrt{5}$]\\
$\Lambda_c\bar{D}^*$&&[$\frac{2}{3}$,0]&[0,$2\sqrt{3}$]&[0,$-4$]&[0,$2\sqrt{2}$]&$\Sigma_c^*\bar{D}$&&[$-\frac{10}{3}$,0]&[0,$-\frac{10}{3\sqrt{3}}$]&[0,$-\frac{10\sqrt{\frac{5}{3}}}{3}$]\\
$\Sigma_c\bar{D}$ &                 &&[$-\frac{10}{3}$,0]  &[0,$-\frac{20}{3\sqrt{3}}$] &[0,$\frac{10\sqrt{\frac{2}{3}}}{3}$]&$\Sigma_c\bar{D}^*$&&&[$-\frac{10}{3}$,$-\frac{20}{9}$]&[0,$\frac{10\sqrt{5}}{9}$]\\
$\Sigma_c\bar{D}^*$&&&&[$-\frac{10}{3}$,$\frac{40}{9}$]&[0,$\frac{10\sqrt{2}}{9}$]&$\Sigma_c^*\bar{D}^*$&&&&[$-\frac{10}{3}$,$\frac{20}{9}$]\\
$\Sigma_c^*\bar{D}^*$&&&&&[$-\frac{10}{3}$,$\frac{50}{9}$]\\
\bottomrule[0.8pt]
\end{tabular}
\end{table*}

\begin{table*}[htbp]
\renewcommand\arraystretch{1.5}
\caption{The matrix elements of
[$\langle\bm{\lambda}_1\cdot\bm{\lambda}_2\rangle$,
$\langle\bm{\lambda}_1\cdot\bm{\lambda}_2\bm{\sigma}_1\cdot\bm{\sigma}_2\rangle$]
for the meson-baryon channels associated with the $J^{P}=1/2^-$ and
$3/2^{-}$ $P_{cs}$ systems.}\label{PcsEF}
\begin{tabular}{cccccccc|ccccccccccccccccc}
\toprule[0.8pt]
\multicolumn{8}{c|}{$\mathbb{V}_{1/2}^{P_{cs}}$}&\multicolumn{6}{c}{$\mathbb{V}_{3/2}^{P_{cs}}$}\\
\hline
Channel& $\Lambda_c\bar{D}_s$&$\Lambda_c\bar{D}_s^*$& $\Xi_c\bar{D}$& $\Xi_c\bar{D}^*$& $\Xi_c^\prime \bar{D}$& $\Xi_c^\prime \bar{D}^*$& $\Xi_c^*\bar{D}^*$ &Channel& $\Lambda_c\bar{D}_s^*$& $\Xi_c\bar{D}^*$& $\Xi_c^* \bar{D}$& $\Xi_c^\prime \bar{D}^*$& $\Xi_c^*\bar{D}^*$ \\
\hline
$\Lambda_c\bar{D}_s$&$[-\frac{4}{3},0]$&[0,0]&$[2\sqrt{2},0]$&$[0,0]$&$[0,0]$&$[0,2\sqrt{2}]$&$[0,4]$&$\Lambda_c\bar{D}_s^*$&$[-\frac{4}{3},0]$&$[2\sqrt{2},0]$&$[0,-2\sqrt{2}]$&$[0,2\sqrt{\frac{2}{3}}]$&$[0,2\sqrt{\frac{10}{3}}]$\\
$\Lambda_c\bar{D}_s^*$&&$[-\frac{4}{3},0]$&$[0,0]$&$[2\sqrt{2},0]$&$[0,2\sqrt{2}]$&$[0,-4\sqrt{\frac{2}{3}}]$&$[0,\frac{4}{\sqrt{3}}]$&$\Xi_c\bar{D}^*$&&$[-\frac{10}{3},0]$&$[0,-2]$&$[0,\frac{2}{\sqrt{3}}]$&$[0,2\sqrt{\frac{5}{3}}]$\\
$\Xi_c\bar{D}$&&&$[-\frac{10}{3},0]$&$[0,0]$&$[0,0]$&$[0,2]$&$[0,2\sqrt{2}]$&$\Xi_c^* \bar{D}$&&&$[-\frac{10}{3},0]$&$[0,-\frac{10}{3\sqrt{3}}]$&$[0,-\frac{10\sqrt{\frac{5}{3}}}{3}]$\\
$\Xi_c\bar{D}^*$&&&&$[-\frac{10}{3},0]$&$[0,2]$&$[0,-\frac{4}{\sqrt{3}}]$&$[0,2\sqrt{\frac{2}{3}}]$&$\Xi_c^\prime \bar{D}^*$&&&&$[-\frac{10}{3},-\frac{20}{9}]$&$[0,\frac{10\sqrt{5}}{9}]$\\
$\Xi_c^\prime \bar{D}$&&&&&$[-\frac{10}{3},0]$&$[0,-\frac{20}{3\sqrt{3}}]$&$[0,\frac{10\sqrt{\frac{2}{3}}}{3}]$&$\Xi_c^*\bar{D}^*$&&&&&$[-\frac{10}{3},\frac{20}{9}]$\\
$\Xi_c^\prime \bar{D}^*$&&&&&&$[-\frac{10}{3},\frac{40}{9}]$&$[0,\frac{10\sqrt{2}}{9}]$\\
$\Xi_c^*\bar{D}^*$&&&&&&&$[-\frac{10}{3},\frac{50}{9}]$\\
\bottomrule[0.8pt]
\end{tabular}
\end{table*}

We can directly obtain the effective potentials associated with the
$P_c$ and $P_{cs}$ states from Tables \ref{PcEF} and \ref{PcsEF},
respectively. For example, the explicit form of the effective
potential matrix for the $J=3/2$ $P_c$ states is
\begin{eqnarray}
\mathbb{V}_{3/2}^{P_c}&&\nonumber\\
=&&\left(
  \begin{array}{ccccc}
    \frac{2}{3}\tilde{g}_s&-2\sqrt{3}\tilde{g}_a&2\tilde{g}_a&2\sqrt{5}\tilde{g}_a\\
    -2\sqrt{3}\tilde{g}_a&-\frac{10}{3}\tilde{g}_s&-\frac{10}{3\sqrt{3}}\tilde{g}_a&-\frac{10\sqrt{\frac{5}{3}}}{3}\tilde{g}_a\\
    2\tilde{g}_a&-\frac{10}{3\sqrt{3}}\tilde{g}_a&-\frac{10}{3}\tilde{g}_s-\frac{20}{9}\tilde{g}_a&\frac{10\sqrt{5}}{9}\tilde{g}_a\\
    2\sqrt{5}\tilde{g}_a&-\frac{10\sqrt{\frac{5}{3}}}{3}\tilde{g}_a&\frac{10\sqrt{5}}{9}\tilde{g}_a&-\frac{10}{3}\tilde{g}_s+\frac{20}{9}\tilde{g}_a\\
  \end{array}
\right).\nonumber\\
\end{eqnarray}
Similarly, the effective potential matrixes
$\mathbb{V}_{1/2}^{P_{c}}$, $\mathbb{V}_{1/2}^{P_{cs}}$, and
$\mathbb{V}_{3/2}^{P_{cs}}$ can also be obtained directly from Table
\ref{PcEF} and \ref{PcsEF}.

We use the masses of the observed $P_c$ states as input to determine
the coupling constants $\tilde{g}_s$ and $\tilde{g}_a$. In our
previous work, we find that the Lagrangian in Eq. (\ref{lag}) can
give a satisfactory description of the observed $T_{cc}$
\cite{LHCb:2021auc,LHCb:2021vvq}, $P_c$, and $P_{cs}$ states if we
assign the $P_{c}(4440)$ and $P_{c}(4457)$ as the
$I(J^P)=1/2(1/2^-)$ and $1/2(3/2^-)$ states. For consistency, we
still adopt this set of assignments and use the masses of the
$P_c(4440)$ and $P_{c}(4457)$ as inputs. In the coupled-channel
formalism, the bound/quasi-bound states in the $J^P=1/2^-$ and
$3/2^-$ $P_{c}$ systems satisfy the following equations
\begin{eqnarray}
\text{Re}\left|\left|\bm{1}-\mathbb{V}_{1/2}^{P_c}\mathbb{G}_{1/2}^{P_c}\right|\right|=0,\label{M1}\\
\text{Im}\left|\left|\bm{1}-\mathbb{V}_{1/2}^{P_c}\mathbb{G}_{1/2}^{P_c}\right|\right|=0,\label{M2}\\
\text{Re}\left|\left|\bm{1}-\mathbb{V}_{3/2}^{P_c}\mathbb{G}_{3/2}^{P_c}\right|\right|=0,\label{M3}\\
\text{Im}\left|\left|\bm{1}-\mathbb{V}_{3/2}^{P_c}\mathbb{G}_{3/2}^{P_c}\right|\right|=0.\label{M4}
\end{eqnarray}
These four equations can be solved numerically and we get
\begin{eqnarray}
\tilde{g}_s=8.28\quad\text{GeV}^{-2},\quad
\tilde{g}_a=-1.46\quad\text{GeV}^{-2}.
\end{eqnarray}
The imaginary part of the pole positions of the $P_c(4440)$ and
$P_{c}(4457)$ can also be obtained from Eqs. (\ref{M1}-\ref{M4}).

\subsection{Flavor-spin symmetry of the $P_c$ and $P_{cs}$ systems in the single-channel formalism}

With the determined parameters $\tilde{g}_s$ and $\tilde{g}_a$, we
first present our single-channel results for the considered $P_c$
and $P_{cs}$ systems and demonstrate that we can relate the $P_c$
and $P_{cs}$ systems from their interactions constrained by the
SU(3) and heavy quark symmetries.

Although the $\Xi_c\bar{D}$ and $\Sigma_c\bar{D}$ belong to
different multiplets, in Ref. \cite{Chen:2021spf} we proposed that
there exists a generalized flavor-spin symmetry between two-body
heavy-flavor systems. For two different heavy-flavor meson-baryon
systems, if they both possess the same flavor ($\langle
H^f_1H^f_2|\bm{\lambda}_1\cdot\bm{\lambda}_2|H^f_1H^f_2\rangle$) and
spin ($\langle
H^s_1H^s_2|\bm{\sigma}_1\cdot\bm{\sigma}_2|H^s_1H^s_2\rangle$)
matrix elements, they will still have identical effective potentials
in the SU(3) and heavy quark limits.

In the single-channel formalism, we present the masses and binding
energies of the $P_c$ and $P_{cs}$ states in Table \ref{SCPcPcs}.
The theoretical uncertainties are introduced by considering the
experimental errors of the masses of the $P_{c}(4440)$ and
$P_{c}(4457)$. We collect the $P_c$ and $P_{cs}$ states that share
identical effective potentials in the same row. As listed in Table
\ref{SCPcPcs}, the $P_c$ and $P_{cs}$ states have similar binding
energies in the same row and can be related via a flavor-spin
symmetry.

\begin{table*}[htbp]
\renewcommand\arraystretch{1.5}
\caption{In the single-channel formalism, the binding energies of
the $P_c$ and $P_{cs}$ states that share the same effective
potentials in the SU(3) and heavy quark limits. All results are in
units of MeV.}\label{SCPcPcs}
\begin{tabular}{ccccccccccccccc}
\toprule[0.8pt]
$P_c$&Mass&BE&$P_{cs}$&Mass&BE&$V$\\
\hline
\multirow{2}{*}{$[\Sigma_c\bar{D}]^{\frac{1}{2}}$}&\multirow{2}{*}{$4312.7^{+4.1}_{-2.6}$}&\multirow{2}{*}{$-8.1^{+4.1}_{-2.6}$}&$[\Xi_c\bar{D}]^{\frac{1}{2}}$&$4328.5^{+4.1}_{-2.7}$&$-8.2^{+4.1}_{-2.7}$&\multirow{4}{*}{$-\frac{10}{3}\tilde{g}_s$}&\\
&&&$[\Xi_c\bar{D}^*]^{\frac{1}{2},\frac{3}{2}}$&$4468.3^{+4.5}_{-2.9}$&$-9.7^{+4.5}_{-2.9}$\\
\multirow{2}{*}{$[\Sigma_c^*\bar{D}]^{\frac{3}{2}}$}&\multirow{2}{*}{$4376.9^{+4.2}_{-2.7}$}&\multirow{2}{*}{$-8.5^{+4.2}_{-2.7}$}&$[\Xi_c^\prime\bar{D}]^{\frac{1}{2}}$&$4437.2^{+4.5}_{-2.8}$&$-8.8^{+4.3}_{-2.8}$\\
&&&$[\Xi_c^*\bar{D}]^{\frac{3}{2}}$&$4503.9^{+4.4}_{-2.8}$&$-9.3^{+4.4}_{-2.8}$\\
\hline
$[\Sigma_c\bar{D}^*]^{\frac{1}{2}}$&$4438.9^{+4.9}_{-8.9}$&$-23.2^{+4.9}_{-8.9}$&$[\Xi_c^\prime\bar{D}^*]^{\frac{1}{2}}$&$4562.9^{+2.8}_{-9.1}$&$-24.5^{+2.8}_{-9.1}$&$-\frac{10}{3}\tilde{g}_s+\frac{40}{9}\tilde{g}_a$\\
$[\Sigma_c\bar{D}^*]^{\frac{3}{2}}$&$4457.5^{+3.7}_{-1.8}$&$-4.6^{+3.7}_{-1.8}$&$[\Xi_c^\prime\bar{D}^*]^{\frac{3}{2}}$&$4582.2^{+4.0}_{-2.0}$&$-5.2^{+4.0}_{-2.0}$&$-\frac{10}{3}\tilde{g}_s-\frac{20}{9}\tilde{g}_a$\\
$[\Sigma_c^*\bar{D}^*]^{\frac{1}{2}}$&$4498.8^{+6.6}_{-6.0}$&$-27.9^{+6.6}_{-6.0}$&$[\Xi_c^*\bar{D}^*]^{\frac{1}{2}}$&$4625.3^{+6.8}_{-12.7}$&$-29.2^{+6.8}_{-12.7}$&$-\frac{10}{3}\tilde{g}_s+\frac{50}{9}\tilde{g}_a$\\
$[\Sigma_c^*\bar{D}^*]^{\frac{3}{2}}$&$4510.3^{+4.1}_{-4.1}$&$-16.4^{+4.1}_{-4.1}$&$[\Xi_c^*\bar{D}^*]^{\frac{3}{2}}$&$4637.9^{+4.3}_{-4.2}$&$-16.6^{+4.3}_{-4.2}$&$-\frac{10}{3}\tilde{g}_s+\frac{20}{9}\tilde{g}_a$\\
\bottomrule[0.8pt]
\end{tabular}
\end{table*}

\subsection{The masses of $P_c$ states in the multi-channel formalism}

Then we explore how the coupled-channel effect influences the masses
of the $P_c$ states. As can be seen from Eq. (\ref{op}), the
effective potential consists of two parts, i.e., the central term
($\tilde{g}_s\bm{\lambda}_1\cdot\bm{\lambda}_2$) and the spin-spin
interaction
($\tilde{g}_a\bm{\lambda}_1\cdot\bm{\lambda}_2\bm{\sigma}_1\cdot\bm{\sigma}_2$)
term. Since the determined $\tilde{g}_s$ is much larger than
$\tilde{g}_a$, the central term dominates the total effective
potential, and therefore determines whether the considered system
can form a bound state.

As given in Table \ref{PcEF}, for the matrix elements in the $P_c$
system, all the diagonal matrix elements have central terms, and
some of them have corrections from the spin-spin interaction terms.
The off-diagonal terms only consist of the spin-spin interaction
terms. Thus, before we perform a practical multi-channel calculation
of the $P_c$ system, we may anticipate that the coupled-channel
effect would have small corrections to the masses of the $P_c$
states.

As discussed in Sec. \ref{sec2}, we include five and four channels
to study the $J=1/2$ and $J=3/2$ $P_c$ states, respectively. For the
$J=1/2$ channels, according to their thresholds, we consider five
energy regions
\begin{eqnarray}
E&\leq&m_{\Lambda_c\bar{D}},\\
m_{\Lambda_c\bar{D}}<&E&\leq m_{\Lambda_c\bar{D}^*},\\
m_{\Lambda_c\bar{D}^*}<&E&\leq m_{\Sigma_c\bar{D}},\\
m_{\Sigma_c\bar{D}}<&E&\leq m_{\Sigma_c\bar{D}^*},\\
m_{\Sigma_c\bar{D}^*}<&E&\leq m_{\Sigma_c^*\bar{D}^*}.
\end{eqnarray}
We search the bound (quasi-bound) state solutions below the higher
threshold in each energy region on the first Riemann sheet. The
bound (quasi-bound) state solutions of the $J=3/2$ $P_c$, $J=1/2$
and $3/2$ $P_{cs}$ states can be found by repeating the same
procedure. We present the obtained $P_c$ states in Table \ref{Pc
multi}. We do not find any bound states below the $\Lambda_c\bar{D}$
threshold. Thus, all the obtained resonances ($E_R$) listed in Table
\ref{Pc multi} should refer to quasi-bound states and have imaginary
parts ($\text{Im}(E_R)$). Since we only include the two body
open-charm decay channels, the estimated widths ($\Gamma$) in Table
\ref{Pc multi} are smaller than experimental widths. By comparing
the masses of $P_c$ states in Table \ref{SCPcPcs} and \ref{Pc
multi}, we find that the coupled-channel effect indeed have small
influences to the masses of the $P_c$ states.

\begin{table}[!htbp]
\centering
\renewcommand\arraystretch{1.5}
\caption{The results of $P_c$ states obtained in the coupled-channel
formalism. Here, $\Gamma=-2\text{Im}(E_R)$, all the results are in
units of MeV.}\label{Pc multi}
\begin{tabular}{ccccccccccccccc}
\toprule[0.8pt]
&&\multicolumn{2}{c}{Our}&\multicolumn{2}{c}{Exp}\\
\hline
State&$J^P$&Mass &$\Gamma$&Mass&Width\\
\hline
$P_c(4312)$&$\frac{1}{2}^-$&$4308.2^{+2.6}_{-4.5}$&$2.6^{+2.4}_{-1.7}$&$4311.9_{-0.9}^{+7.0}$&$10\pm5$\\
$P_c(4440)$&$\frac{1}{2}^-$&$4440.3^{+4.0}_{-5.0}$ (input)&$9.8^{+4.6}_{-5.8}$&$4440.3^{+4.0}_{-5.0}$&$21^{+10}_{-11}$\\
$P_c(4457)$&$\frac{3}{2}^-$&$4457.7^{+4.0}_{-1.8}$ (input)&$2.0^{+1.4}_{-0.8}$&$4457.3^{+4.0}_{-1.8}$&$6.4^{+6.0}_{-2.8}$\\
\hline
$P_c(4380)$&$\frac{3}{2}^-$&$4373.3^{+3.4}_{-6.8}$&$5.2^{+20.2}_{-3.5}$&$-$&$-$\\
$P_c(4500)$&$\frac{1}{2}^-$&$4501.4^{+5.0}_{-6.2}$&$8.8^{+17.2}_{-5.4}$&$-$&$-$\\
$P_c(4510)$&$\frac{3}{2}^-$&$4513.4^{+5.8}_{-3.1}$&$7.6^{+9.4}_{-0.0}$&$-$&$-$\\
\bottomrule[0.8pt]
\end{tabular}
\end{table}

\subsection{A numerical experiment on the ($\Lambda_c\bar{D}^{(*)}_s$, $\Xi_c\bar{D}^{(*)}$) coupled-channel systems}

There exists an important difference between the effective potential
matrices in the $P_c$ and $P_{cs}$ systems. As presented in Tables
\ref{PcEF} and \ref{PcsEF}, the diagonal matrix elements in the
$P_{cs}$ system are very similar to those of the $P_c$ system. But
for the off-diagonal matrix elements, the effective potentials of
the $\Lambda_c\bar{D}_s-\Xi_c\bar{D}$ and
$\Lambda_c\bar{D}_s^*-\Xi_c\bar{D}^*$ channels in the $P_{cs}$
system with $J=1/2$ or $3/2$ consist of central terms. These terms
may give considerable corrections to the spectrum of the $P_{cs}$
states.

For the $J=1/2$ and $J=3/2$ $P_{cs}$ systems, as given in Eq.
(\ref{Pcs12}-\ref{Pcs32}) we need to perform seven and five
coupled-channel calculations. Before we perform such complete
calculations, we first perform a detailed discussion on the
($\Lambda_c\bar{D}_s$, $\Xi_c\bar{D}$) and ($\Lambda_c\bar{D}_s^*$,
$\Xi_c\bar{D}^*$) coupled-channel systems.

The effective potential matrixes of the $J=1/2$
($\Lambda_c\bar{D}_s$, $\Xi_c\bar{D}$), ($\Lambda_c\bar{D}_s^*$,
$\Xi_c\bar{D}^*$) systems, and the $J=3/2$ ($\Lambda_c\bar{D}_s^*$,
$\Xi_c\bar{D}^*$) system share the same expressions in the heavy
quark limit. From Table \ref{PcsEF}, we obtain the corresponding
effective potential matrix
\begin{eqnarray}
\mathbb{V}=\left(
             \begin{array}{cc}
               v_{11} & v_{12} \\
               v_{21} & v_{22} \\
             \end{array}
           \right)
\end{eqnarray}
with
\begin{eqnarray}
v_{11}&=&-\frac{4}{3}\tilde{g}_s, \quad v_{22}=-\frac{10}{3}\tilde{g}_s,\\
v_{12}&=&v_{21}=2\sqrt{2}\tilde{g}_s g_x.
\end{eqnarray}
Here, for the diagonal matrix elements listed in Table \ref{PcEF}
and \ref{PcsEF}, their dominant components are from the exchange of
the non-strange light scalar meson currents. Since the interactions
of the off-diagonal channel
$\Lambda_c\bar{D}^{(*)}_s-\Xi_c\bar{D}^{(*)}$ are introduced via the
exchange of the strange scalar meson currents, we further introduce
a factor $g_x$ to estimate the SU(3) breaking effects. Compared with
the exchange of the non-strange light scalar meson currents, the
off-diagonal matrix elements should be suppressed by the mass of
strange mesons. Thus, we assume $0\leq gx\leq1$. This factor also
reflects the coupling strength of the
$\Lambda_c\bar{D}_s-\Xi_c\bar{D}$ channel. With $g_x=0$, the
$\Lambda_c\bar{D}^{(*)}_s$ does not couple to the $\Xi_c\bar{D}^{(*)}$
channel. With $g_x=1.0$, the $\Lambda_c\bar{D}^{(*)}_s$ couples to the
$\Xi_c\bar{D}^{(*)}$ channel and its coupling strength is set to be
the value in the SU(3) limit.

\begin{figure}[htbp]
    \centering
    \includegraphics[width=1.0\linewidth]{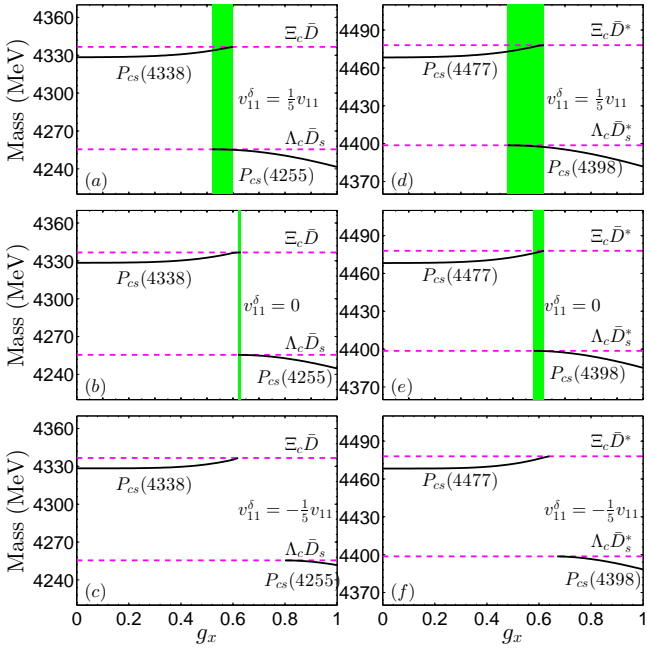}
    \caption{The variations of the masses for the possible bound states in the ($\Lambda_c\bar{D}^{(*)}_s$ $\Xi_c\bar{D}^{(*)}$) two-channel system as the parameter $g_x$ increases. We use the blue-dotted lines to denote the $\Lambda_c\bar{D}^{(*)}_s$ and $\Xi_c\bar{D}^{(*)}$ thresholds. The masses of the $P_{cs}$ states are denoted with black lines. Figs. (a), (d), Figs. (b), (e), and Figs. (c), (f) are obtained at $v_{11}^{\delta}=1/5 v_{11}$, 0, and $-1/5v_{11}$, respectively. The green bands in  Figs. (a), (d), (b), (e) denote that the bound states near the $\Lambda_c\bar{D}_s$ ($\Lambda_c\bar{D}_s^{*}$) and $\Xi_c\bar{D}$ ($\Xi_c\bar{D}^{*}$) thresholds can coexist in this $g_x$ region.}
    \label{LXCC}
\end{figure}

In Fig. \ref{LXCC} (b), we present the variation of the masses for
the bound states $P_{cs}(4338)$ and $P_{cs}(4255)$ as the parameter
$g_x$ increases. The masses of the $P_{cs}(4338)$ and $P_{cs}(4255)$
are denoted with black lines. At $g_x=0$, the $\Lambda_c\bar{D}_s$
channel itself has a weak attractive force $v_{11}=-4/3\tilde{g}_s$,
and this force is too weak to form a $\Lambda_c\bar{D}_s$ bound
state. On the contrary, the $\Xi_c\bar{D}$ channel can form a bound
state and its mass is about $M=4329$ MeV, slightly smaller than the
experimental value. As the $g_x$ increases, the attractive force of
the $P_{cs}(4338)$ decreases and its mass moves closer to the
$\Xi_c\bar{D}$ threshold. In a very narrow region $0.62\leq g_x\leq
0.64$, the attractive force is just enough to form a $P_{cs}(4338)$
bound state at the $\Xi_c\bar{D}$ threshold and the weak attractive
channel $\Lambda_c\bar{D}_s$ starts to form a bound state due to the
$\Xi_c\bar{D}-\Lambda_c\bar{D}_s$ coupling. Only in this very narrow
region, the $P_{cs}(4338)$ and $P_{cs}(4255)$ can coexist as
quasi-bound states. At $g_x>0.64$, the
$\Xi_c\bar{D}-\Lambda_c\bar{D}_s$ coupling further weakens the
attractive force of the $\Xi_c\bar{D}$ channel and the
$P_{cs}(4338)$ no longer exists as a quasi-bound state, while the
attractive force of the $\Lambda_c\bar{D}_s$ channel becomes
stronger and its mass will decrease. The observation of the
$P_{cs}(4338)$ by LHCb seems to exclude the parameter region
$0.64<g_x<1.0$.

Here, we also check the pole position of the $P_{cs}(4338)$ at
$g_x>0.64$ in the energy region slightly above the $\Xi_c\bar{D}$
threshold. We find that the pole of the $P_{cs}(4338)$ still exists
in the first Riemann sheet. This is mainly due to the fact that the
$\Lambda_c\bar{D}_s-\Xi_c\bar{D}$ coupling leads the $P_{cs}(4338)$
to be a state that have a considerable width, thus the central value
of the $P_{cs}(4338)$ mass may cross the $\Xi_c\bar{D}$ threshold.
In this case, the $P_{cs}(4338)$ should be interpreted as a
quasi-bound state above the $\Xi_c\bar{D}$ threshold. Nevertheless,
in this work, we restrict our scope to the case that the masses of
the bound/quasi-bound states are below their corresponding
thresholds.

To understand why the $g_x$ region that allows the $P_{cs}(4338)$
and $P_{cs}(4255)$ states to coexist is so narrow, we further check
the role of the $\Lambda_c\bar{D}_s$ channel in our two-channel
model. We allow the effective potential of the $\Lambda_c\bar{D}_s$
channel to have a $20\%$ shift, i.e.,
\begin{eqnarray}
v^\prime_{11}=v_{11}+v^{\delta}_{11},\quad
v^{\delta}_{11}=0,\pm\frac{1}{5}v_{11},
\end{eqnarray}
and further check how the masses of the $P_{cs}(4338)$ and
$P_{cs}(4255)$ change as we increase the $g_x$.

The channel $\Lambda_c\bar{D}_s$ itself has a weak attractive force,
as presented in Fig. \ref{LXCC} (a), at $g_x=0$. After we increase
this force by 20$\%$, this single-channel still can not form a bound
state. But the $g_x$ region that allows these two $P_{cs}$ states to
coexist becomes broader. On the contrary, as illustrated in Fig.
\ref{LXCC} (c), if we decrease the attractive force of the
$\Lambda_c\bar{D}_s$ channel by 20$\%$, the $P_{cs}(4338)$ and
$P_{cs}(4255)$ can not coexist no matter how we adjust the
off-diagonal $\Lambda_c\bar{D}_s-\Xi_c\bar{D}$ coupling. Thus, the
narrow $g_x$ region that the $P_{cs}(4338)$ and $P_{cs}(4255)$ can
coexist is due to the fact that the $\Lambda_c\bar{D}_s$ channel has
a small but non-negligible attractive force.

The results for the $J=1/2$ and $3/2$
($\Lambda_c\bar{D}_s^*$,$\Xi_c\bar{D}^*$) coupled-channels are
presented in Fig. \ref{LXCC} (d-f). We find that the roles of the
predicted $P_{cs}(4477)$ and $P_{cs}(4398)$ with $J^P=1/2^-$ or
$3/2^-$ are very similar to those of the $P_{cs}(4338)$ and
$P_{cs}(4255)$ with $J^{P}=1/2^-$, respectively.

\subsection{The results of $P_{cs}$ system in the coupled-channel formalism}
\begin{figure}[htbp]
    \centering
    \includegraphics[width=1.0\linewidth]{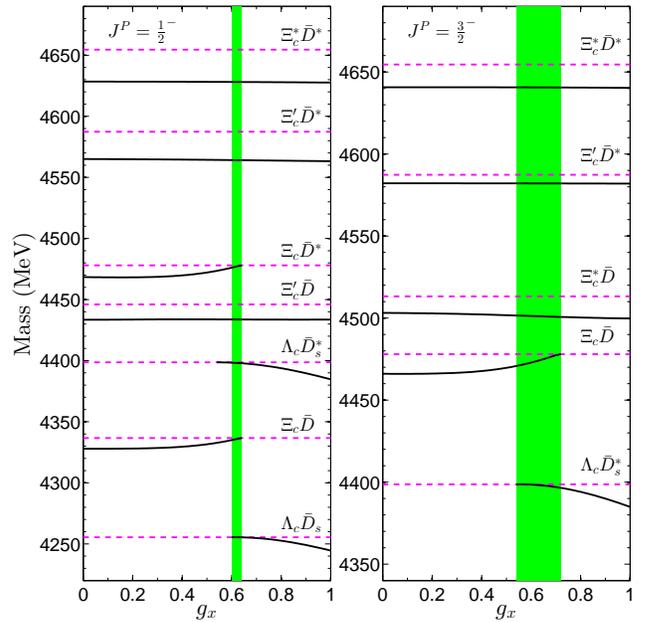}
    \caption{The variations of the masses for the $P_{cs}$ states with $J^{P}=1/2^-$ and $J^P=3/2^-$ as the $g_x$ increases. We use the blue-dotted lines to denote the considered meson-baryon thresholds. The masses of the obtain bound states are denoted with black lines. The green bands denote that the bound states near the $\Lambda_c\bar{D}_s^{(*)}$ or $\Xi_c\bar{D}^{(*)}$ thresholds can coexist in this $g_x$ region.}
    \label{mulcc}
\end{figure}

We present our complete multi-channel calculations on the $J=1/2$
and $J=3/2$ $P_{cs}$ systems in Fig. \ref{mulcc}. We find that only
the bound states close to the $\Lambda_c\bar{D}_s^{(*)}$ and
$\Xi_c\bar{D}^{(*)}$ channels have significant dependence on the
$g_x$, since these bound states can couple to the
$\Lambda_c\bar{D}_s^{(*)}$ and $\Xi_c\bar{D}^{(*)}$ channels through
non-negligible central terms, while the bound states that can only
couple to the $\Lambda_c\bar{D}_s^{(*)}$ and $\Xi_c\bar{D}^{(*)}$
channels via the spin-spin interaction terms have very tiny
dependence on the parameter $g_x$.

To further present our numerical results, we fix the parameter $g_x$
at 0.5 and 0.62. We denote these two cases as case 1 and case 2. The
case 1 and case 2 correspond to the results that the possible
$P_{cs}(4255)$ signal does not/does exist. The results of these two
cases are listed in Table \ref{mulPcs}.
\begin{table*}[htbp]
\renewcommand\arraystretch{1.5}
\caption{The results of the $P_{cs}$ states calculated at $g_x=0.50$
and $g_x=0.62$ in the coupled-channel formalism. All results are in
units of MeV.}\label{mulPcs}
\begin{tabular}{c|ccc|ccccccc}
\toprule[0.8pt]
&\multicolumn{3}{c}{$g_x=0.50$}&\multicolumn{3}{c}{$g_x=0.62$}\\
\hline
States&Mass&$\Gamma$&BE&Mass&$\Gamma$&BE\\
$[\Lambda_c\bar{D}_s]^{\frac{1}{2}}$&$-$&$-$&$-$&$4255.5_{-0.7}^{+0.0}$&0.0&$-0.0_{-0.7}^{+0.0}$\\
$[\Lambda_c\bar{D}_s^*]^{\frac{1}{2}}$&$-$&$-$&$-$&$4398.1^{+0.2}_{-1.5}$&0.0&$-0.6^{+0.2}_{-1.5}$\\
$[\Lambda_c\bar{D}_s^*]^{\frac{3}{2}}$&$-$&$-$&$-$&$4398.3^{+0.4}_{-1.3}$&$0.0$&$-0.4^{+0.4}_{-1.3}$\\
$[\Xi_c\bar{D}]^{\frac{1}{2}}$&$4331.6_{-1.5}^{+2.6}$&$17.8_{-7.6}^{+3.6}$&$-5.0^{+2.6}_{-1.5}$&$4335.9_{-2.1}^{+0.7}$&$26.0^{+5.0}_{-10.2}$&$-0.7_{-2.1}^{+0.7}$\\
$[\Xi_c\bar{D}^*]^{\frac{1}{2}}$&$4472.1^{+2.5}_{-1.6}$&$23.4^{+5.2}_{-6.0}$&$-5.9^{+2.5}_{-1.6}$&$4477.1^{+0.5}_{-0.8}$&$33.0^{+6.0}_{-8.0}$&$-0.9^{+0.5}_{-0.8}$\\
$[\Xi_c\bar{D}^*]^{\frac{3}{2}}$&$4469.7^{+1.9}_{-6.6}$&$14.6^{+4.4}_{-10.6}$&$-8.3^{+1.9}_{-6.6}$&$4473.7^{+2.7}_{-8.7}$&$20.2^{+5.8}_{-13.6}$&$-4.7^{+2.7}_{-8.7}$\\
\hline \hline
$[\Xi_c^\prime\bar{D}]^{\frac{1}{2}}$&$4433.8^{+3.4}_{-4.8}$&$0.8_{-0.6}^{+5.0}$&$-12.2^{+3.4}_{-4.8}$&$4433.7_{-5.2}^{+3.5}$&$0.4_{-0.0}^{+2.6}$&$-12.3^{+3.5}_{-5.2}$\\
$[\Xi_c^*\bar{D}]^{\frac{3}{2}}$&$4501.8^{+4.4}_{-3.4}$&$7.7^{+22.3}_{-5.1}$&$-11.4^{+4.4}_{-3.4}$&$4501.2^{+4.2}_{-3.7}$&$7.8^{+25.4}_{-5.3}$&$-12.0^{+4.2}_{-3.7}$\\
$[\Xi_c^\prime\bar{D}^*]^{\frac{1}{2}}$&$4564.4^{+3.4}_{-4.0}$&$4.5^{+9.5}_{-2.7}$&$-23.0^{+3.4}_{-4.0}$&$4564.1_{-7.4}^{+3.3}$&$4.8^{+9.2}_{-2.8}$&$-23.3_{-7.4}^{+3.3}$\\
$[\Xi_c^\prime\bar{D}^*]^{\frac{3}{2}}$&$4582.1^{+4.3}_{-2.0}$&$1.9_{-1.1}^{+0.7}$&$-5.3_{-2.0}^{+4.3}$&$4582.1^{+4.2}_{-2.0}$&$2.0^{+1.3}_{-1.2}$&$-5.3^{+4.2}_{-2.0}$\\
$[\Xi_c^*\bar{D}^*]^{\frac{1}{2}}$&$4628.2_{-7.1}^{+3.5}$&$5.0_{-3.0}^{+9.0}$&$-26.3_{-7.1}^{+3.5}$&$4628.1^{+5.1}_{-7.3}$&$5.2_{-3.2}^{+24.5}$&$-26.4_{-7.3}^{+5.1}$\\
$[\Xi_c^*\bar{D}^*]^{\frac{3}{2}}$&$4640.7^{+5.8}_{-3.2}$&$7.0_{-0.8}^{+8.2}$&$-13.8^{+5.8}_{-3.2}$&$4640.6_{-3.2}^{+5.9}$&$7.2^{+8.4}_{-4.4}$&$-13.9^{+5.9}_{-3.2}$\\
\bottomrule[0.8pt]
\end{tabular}
\end{table*}

Comparing the masses of the $P_{cs}$ states calculated in the
single-channel formalism (Table \ref{SCPcPcs}) with the results
obtained in the coupled-channel formalism (Table \ref{mulPcs}), we
infer that the off-diagonal channels that only consist of the
spin-spin interaction terms have small influence on the masses of
the $P_{cs}$ states, which is very similar to the $P_c$ system. From
Table \ref{mulPcs}, we find that there exist three extra $P_{cs}$
states below the $\Lambda_c\bar{D}^{(*)}_s$ thresholds in the case 2.

For the $P_{cs}(4338)$ state, due to its strong coupling to the
$\Lambda_c\bar{D}_s$ channel, the width of this state is broader
than the result given by LHCb. Note that in our calculation, we only
include the open-charm two body meson-baryon channels. Thus, the
width predicted by our model should be regarded as the lower limit
of the experimental width. Since the $P_{cs}(4338)$ is reported in
the $B\rightarrow J/\Psi\Lambda\bar{p}$ channel, the narrow width of
the $P_{cs}(4338)$ found by the LHCb may be due to the small phase
space of this $B$ meson decay process. Thus, confirming the
$P_{cs}(4338)$ in other decay processes is important to pin down its
resonance parameters.

Besides, we also find that the $P_{cs}$ states that are close to the
$\Xi_c\bar{D}^{(*)}$ states are broader than the other $P_{cs}$
states due to its strong coupling to the $\Lambda_c\bar{D}_s^{(*)}$
channel. Thus, our results suggest that there exist two $J^P=1/2^-$
and $J^{P}=3/2^-$ quasi-bound states near the $\Xi_c\bar{D}^*$
region. This region is close to the reported $P_{cs}(4459)$, and the
two-peak structure in this region has been discussed in many
literatures
\cite{Xiao:2021rgp,Wang:2019nvm,Chen:2020uif,Wang:2020eep,Zhu:2021lhd}.
The results from our model provide a new possibility, i.e., the two
$P_{cs}$ structures in this region may have a significant overlap in
the $J/\Psi\Lambda$ invariant spectrum due to their considerable
widths. The decay behaviors of the $P_{cs}(4459)$ have been
discussed in Refs.
\cite{Du:2021bgb,Ortega:2022uyu,Azizi:2021utt,Chen:2021tip,Shen:2019evi}.
The decay widths and decay patterns are valuable in identifying the
structure of the $P_{cs}(4459)$. Further investigations on the total
and partial decay widths will be crucial to accomplish a thorough
understanding on the $P_{c}$ and $P_{cs}$ states.

\subsection{The correspondence between the $P_c$ and $P_{cs}$ systems}
\begin{figure*}[htbp]
    \centering
    \includegraphics[width=1.0\linewidth]{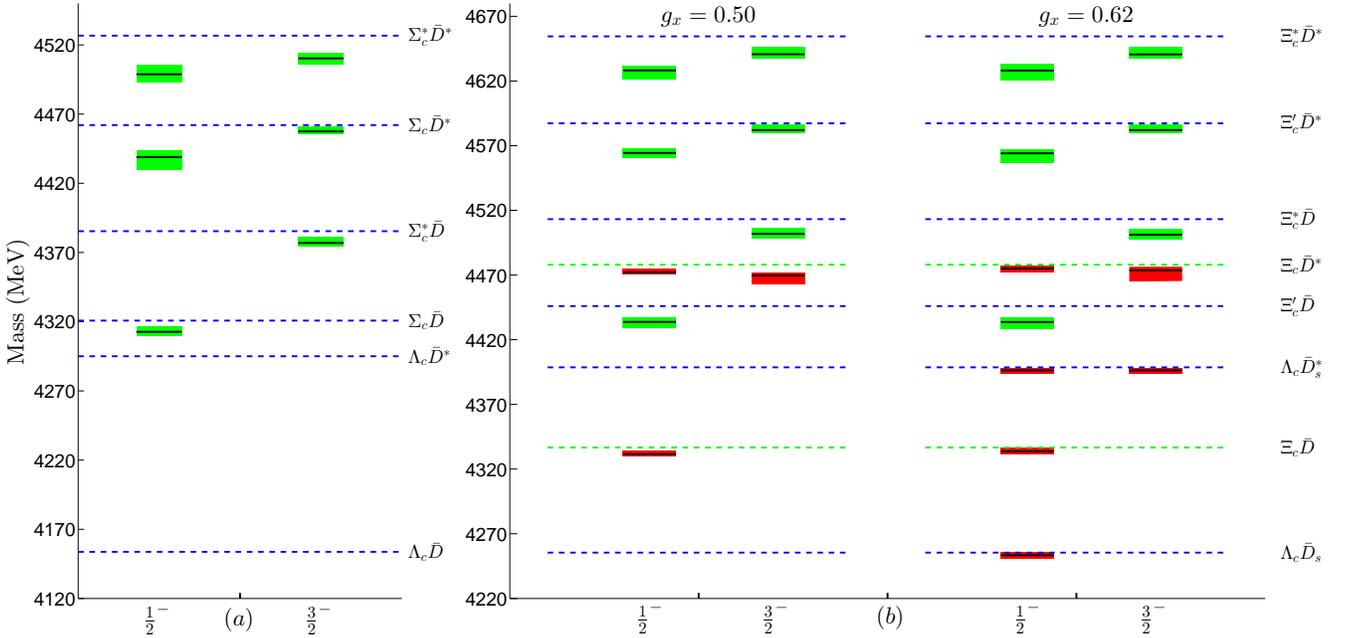}
\caption{The mass spectra of the $P_c$ and $P_{cs}$ states in the
multi-channel formalism, we present the results of $P_{cs}$ system
at $g_x=0.50$ and $g_x=0.62$. The meson-baryon thresholds
$m_{\Lambda_c\bar{D}^{(*)}}$
$m_{\Lambda_c\bar{D}_s^{(*)}}$,$m_{\Sigma_c^{(*)}\bar{D}^{(*)}}$, $
m_{\Xi_c^{\prime(*)}\bar{D}^{(*)}}$ are illustrated with the
blue-dotted lines. The two extra thresholds $\Xi_c\bar{D}$ and
$\Xi_c\bar{D}^*$ in the $P_{cs}$ system are denoted with the
green-dotted lines. We use the black lines to denote the central
values of the obtained $P_c$ and $P_{cs}$ states, their
uncertainties are illustrated with the green and red rectangles. The
six green states in the $P_c$ system can directly correspond to the
six green states in the $P_{cs}$ system.}
    \label{mulPcPcs}
\end{figure*}

Finally, we compare the masses of the $P_c$ and $P_{cs}$ states
obtained from our multi-channel model. Since the mass of constituent
$s$ quark is heavier than that of the $u$, $d$ quarks by about 100
MeV. Thus, we shift the mass plot of the $P_{cs}$ system by 100 MeV
to check the similarities between the $P_{c}$ and $P_{cs}$ states.
We present the multi-channel results for the $P_c$ system in Fig.
\ref{mulPcPcs} (a), and the multi-channel results for the $P_{cs}$
system calculated at $g_x=0.5$ and $g_x=0.62$ are given in Fig.
\ref{mulPcPcs} (b). As can be seen from Fig. \ref{mulPcPcs} (a) and
(b), the meson-baryon thresholds in the $P_c$ and $P_{cs}$ systems
have the following analogies
\begin{eqnarray}
m_{\Lambda_c\bar{D}^{(*)}}&\leftrightarrow& m_{\Lambda_c\bar{D}_s^{(*)}},\\
m_{\Sigma_c^{(*)}\bar{D}^{(*)}}&\leftrightarrow&
m_{\Xi_c^{\prime(*)}\bar{D}^{(*)}}.
\end{eqnarray}
We denote these thresholds with the blue-dotted lines in Fig.
\ref{mulPcPcs}. Besides, there exist two extra meson-baryon
thresholds $\Xi_c\bar{D}$ and $\Xi_c\bar{D}^*$ in the $P_{cs}$
system. These two channels can not directly correspond to the
meson-baryon channels in the $P_c$ system. We denote these two
thresholds with the green-dotted lines.

As can be seen from Fig. \ref{mulPcPcs}, there exist six $P_c$
states with $J^P=1/2^-$ or $3/2^-$. These six states can correspond
to the six states in the $P_{cs}$ system. We denote the masses of
the central values of these 12 states with black lines and their
uncertainties are denoted with green rectangles. According to Fig.
\ref{mulPcPcs}, the experimentally observed $P_c$ states and the
predicted $P_{cs}$ states should have the following analogies
\begin{eqnarray}
P_{c}(4312)&\leftrightarrow& P_{cs}(4434),\\
P_{c}(4440)&\leftrightarrow& P_{cs}(4564),\\
P_{c}(4457)&\leftrightarrow& P_{cs}(4582).
\end{eqnarray}
As indicated in Fig. \ref{mulPcPcs}, if we replace the
$\Lambda_c\bar{D}_s^*$ and $\Xi_c^\prime\bar{D}$ channels with the
$\Xi_c\bar{D}$ and $\Xi_c\bar{D}^*$ channels, respectively, we can
reluctantly obtain the following analogies
\begin{eqnarray}
P_{c}(4312)&\leftrightarrow& P_{cs}(4472),\\
P_{c}(4440)&\leftrightarrow& P_{cs}(4564),\\
P_{c}(4457)&\leftrightarrow& P_{cs}(4582).
\end{eqnarray}
The predicted $P_{cs}(4472)$ may correspond to the reported
$P_{cs}(4459)$. However, such an analogy indicates a considerable
SU(3) breaking effect. In both sets of analogies, the $P_{cs}(4338)$
can not directly correspond to the lowest $P_c(4312)$ state.

There exist three and six extra $P_{cs}$ states that can not
correspond to the states in the $P_c$ system at $g_x=0.50$ and
$g_x=0.62$, respectively. We denote the masses of the central values
of these states with the black lines and their uncertainties are
denoted with red rectangles. Further experimental explorations on
the $P_{cs}$ system may help us to distinguish which case should be
preferred.

\section{Summary}\label{sec4}

Motivated by the recently discovered $P_{cs}(4338)$ from the LHCb
Collaboration, we have performed a multi-channel calculation of the
$I=1/2$ $P_c$ and $I=0$ $P_{cs}$ systems and presented a comparison
between the interactions of the $P_c$ and $P_{cs}$ states in the
SU(3)$_{\text{f}}$ limit and heavy quark limit.

Unlike the $\bar{c}n-cnn$ ($n=u$, $d$) type meson-baryon channels in
the $P_c$ system, we need to consider two types of channels when we
study the $P_{cs}$ system, i.e., the $\bar{c}n-cns$ and
$\bar{c}s-cnn$ meson-baryon channels. This difference will lead to
extra states in the $P_{cs}$ systems.

The effective potentials of the $P_c$ and $P_{cs}$ states are
collectively obtained via a quark-level Lagrangian, which allows us
to construct the correspondence between the $P_c$ and $P_{cs}$
systems.

We use the masses of the $P_c(4440)$ and $P_c(4457)$ as input to
determine the coupling parameters $\tilde{g}_s$ and $\tilde{g}_a$ in
our model. We first study the masses of the $P_c$ states in the
single-channel and coupled-channel formalisms. Since all the
off-diagonal terms in the effective potential matrices consist of
the spin-spin interaction terms, the coupled-channel effect provides
very small corrections to the masses of the $P_c$ states.

There exists an important difference between the $P_c$ system and
$P_{cs}$ system. In the $P_{cs}$ system, the off-diagonal terms
$\Lambda_c\bar{D}^{(*)}_s-\Xi_c\bar{D}^{(*)}$ in the effective
potential matrices consist of the central terms and will have
considerable corrections to the mass spectrum of the $P_{cs}$
states. To clarify the role of the
$\Lambda_c\bar{D}^{(*)}_s-\Xi_c\bar{D}^{(*)}$ coupling, we have
performed a numerical experiment on the
$(\Lambda_c\bar{D}^{(*)}_s,\Xi_c\bar{D}^{(*)})$ coupled-channel
system. Our results suggest that the mass of the $P_{cs}(4338)$ may
shift very close to the $\Xi_c\bar{D}$ threshold by adjusting the
coupling between the $\Xi_c\bar{D}$ and $\Lambda_c\bar{D}_s$
channels. This coupling may also lead to a $P_{cs}(4255)$ state in a
reasonable $g_x$ region.

Then we present our complete multi-channel calculations of the
$P_{cs}$ systems. Since the $P_{cs}(4255)$ is not confirmed by
experiment, we present our numerical results with $g_x=0.50/0.62$,
corresponding to the case that the $P_{cs}(4255)$ does not/does
exist, respectively. Due to the strong
$\Lambda_c\bar{D}_s-\Xi_c\bar{D}$ couplings, our predicted width of
$P_{cs}(4338)$ is broader than the experimental value. The reported
narrower width may be due to the small phase space of the $B$ meson
decay process. Confirming the $P_{cs}(4338)$ state in other
processes will be helpful to pin down its resonance parameters.
There exist two $P_{cs}$ states with $J^P=1/2^-$ and $J^P=3/2^-$
below the $\Xi_c\bar{D}^*$ threshold. The masses of these two states
are close to the mass of the reported $P_{cs}(4459)$. Due to the
$\Lambda_c\bar{D}^*_s-\Xi_c\bar{D}^*$ coupling, these two states
should have considerable widths and may have significant overlap in
the $J/\Psi\Lambda$ invariant spectrum. Further experimental
exploration would be important to test our predictions.

Finally, we present a complete correspondence between the $P_c$ and
$P_{cs}$ states. The observed $P_c(4312)$, $P_{c}(4440)$, and
$P_{c}(4457)$ do not directly correspond to the observed
$P_{cs}(4338)$ and $P_{cs}(4459)$. It is particularly interesting to
find the SU(3) $P_{cs}$ states that may correspond to the observed
$P_c$ states, and to investigate if such a correspondence does
exist. Further experimental researches on these topics will be
helpful to fulfill a complete picture on the spectra of the $P_c$
and $P_{cs}$ systems.

\section*{Acknowledgments}
This research is supported by the National Science Foundation of
China under Grants No. 11975033, No. 12070131001 and No. 12147168.

\end{document}